# Fuzzification-based Feature Selection for Enhanced Website Content Encryption

Mike Nkongolo[1*]

[1*]Department of Informatics, University of Pretoria, Hatfield 0028, Pretoria, Gauteng, South-Africa.

Corresponding author(s). E-mail(s): mike.wankongolo@up.ac.za;

**Abstract**

We propose a novel approach that utilizes fuzzification theory to perform feature selection on website content for encryption purposes. Our objective is to identify and select the most relevant features from the website by harnessing the principles of fuzzy logic. Fuzzification allows us to transform the crisp website content into fuzzy representations, enabling a more nuanced analysis of their characteristics. By considering the degree of membership of each feature in different fuzzy categories, we can evaluate their importance and relevance for encryption. This approach enables us to prioritize and focus on the features that exhibit higher membership degrees, indicating their significance in the encryption process. By employing fuzzification-based feature selection, we aim to enhance the effectiveness and efficiency of website content encryption, ultimately improving the overall internet security.

**Keywords:** Encryption, web mining, security, information retrieval, computer security, fuzzy logic

## 1 Introduction

The rapid growth of the internet and the increasing reliance on digital platforms have raised concerns about the security and privacy of online information [1]. As websites handle vast amounts of sensitive data, ensuring the confidentiality and integrity of their content has become paramount [1, 2]. Encryption techniques play a crucial role in safeguarding this information from unauthorized access and malicious attacks [3]. However, the effectiveness of encryption heavily relies on the selection of relevant features that capture the essential characteristics of the website content. In this research, we propose a novel approach that leverages fuzzification theory to perform feature selection for website content encryption. The objective of our study is to identify and select the most pertinent features from website content by harnessing the principles of fuzzy logic. Fuzzification offers a means to transform the crisp website content into fuzzy representations, enabling a more comprehensive and nuanced analysis of their underlying characteristics [3]. By considering the degree of membership of each feature in different fuzzy categories, we can evaluate their importance and relevance to the encryption process.

The key advantage of our approach lies in the ability to prioritize and focus on the features that exhibit higher membership degrees in specific fuzzy categories, indicating their significance in the encryption process. By employing fuzzification-based feature selection, we aim to enhance the effectiveness and efficiency of website content encryption, thereby improving the overall security of the internet.

Our research builds upon the foundations of fuzzy logic [4] and its applications in various fields [3, 4], including data analysis and decision-making [5]. Fuzzy logic provides a flexible and intuitive framework to handle uncertainty and imprecision [3], which are inherent in website content due to its diverse nature and evolving context. By exploiting the principles of fuzzy logic, we can better capture the inherent fuzziness and linguistic characteristics of website content, leading to more accurate and context aware feature selection for encryption.

In this paper, we will outline the methodology of our proposed approach, including the fuzzification process, fuzzy category definition, and feature selection algorithm. The contributions of this research lie in providing a novel perspective on feature selection for encryption, leveraging the power of fuzzy logic to improve security measures in the digital realm.

Through this research, we aim to contribute to the advancement of website content encryption techniques by employing fuzzy logic-based feature selection [3]. By ensuring the confidentiality and integrity of website content, we can foster a more secure online environment, thereby safeguarding sensitive information and promoting trust in digital platforms.

## 2   Fuzzification: Transforming Crisp Website Content to Fuzzy Representations

Fuzzification is a process that allows the transformation of crisp website content into fuzzy representations [3]. The Crisp website content refers to the original, unmodified data or information present on a website. It is the raw, precise representation of the content without any fuzziness or uncertainty. Crisp website content typically consists of clear and well-defined elements such as text, images, videos, links, and other media components that make up the website's structure and information.

These elements are usually static and do not incorporate any ambiguity or imprecision in their representation. In the context of fuzzification, crisp website content serves as the input that undergoes the transformation into fuzzy representations to enable a more flexible and nuanced analysis.

Let's consider a crisp website content variable $x_i$ that represents the *i*-th character of the content (Figure 1). The goal is to fuzzify this crisp variable into a fuzzy representation $u_i$.

To achieve this, we define fuzzy sets $F_j$ to represent different fuzzy categories (Figure 1). Each fuzzy set $F_j$ is characterized by a membership function $u_{ij}$, which indicates the degree of membership of character $x_i$ in fuzzy category $F_j$ (Figure 1). Fuzzification assigns membership degrees to characters based on their similarity to the characteristics of each fuzzy category [3].

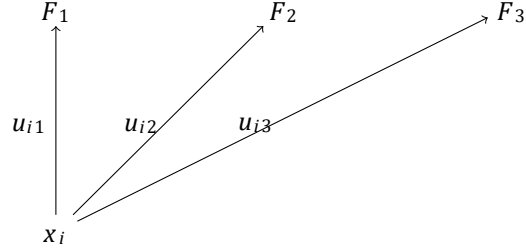

**Fig. 1** Fuzzification process: Assigning membership degrees to characters based on fuzzy categories.

The membership function $u_{ij}$ can be defined using a Gaussian function.

where:

- $x_i$ is the crisp value of the feature, representing the input to the membership function.
- $\mu_j$ is the mean of fuzzy category $F_j$, representing the center of the Gaussian curve.
- $\sigma_j$ is the standard deviation of fuzzy category $F_j$, determining the spread or width of the Gaussian curve.

The membership function $u_{ij}$ calculates the degree of membership of the crisp value $x_i$ in the fuzzy category $F_j$ using a Gaussian curve. The exponent in the equation represents the squared difference between the crisp value and the mean of the fuzzy category. The denominator term $2\sigma_j^2$ controls the spread or width of the Gaussian curve, indicating how much the membership values decrease as the crisp value deviates from the mean.

The exponential function exp raises the base value to the power of the calculated exponent, resulting in a membership value between 0 and 1. A higher membership value indicates a stronger degree of membership of the crisp value in the fuzzy category [3].

This mathematical representation allows us to quantify the similarity of the crisp value to the characteristics of the fuzzy category, enabling fuzzy inference and reasoning in various applications, including fuzzification and fuzzy logic-based systems.

Mathematically, the fuzzification process can be represented as follows:

$$u_{ij} = \text{Membership Function}(x_i, F_j)$$

where $u_{ij}$ represents the degree of membership of character $x_i$ in fuzzy category $F_j$.

By applying appropriate membership functions and fuzzy logic principles, fuzzification allows the crisp website content to be represented in a more flexible and nuanced manner. This transformation enables a more comprehensive analysis of the content's characteristics, capturing the inherent uncertainty and imprecision that often arise in website data.

In the context of website content encryption, fuzzification enables a more sophisticated analysis of the content's features, considering their degree of relevance and importance in the encryption process. By transforming crisp website content into fuzzy representations, fuzzification facilitates the application of encryption techniques that can leverage fuzzy logic principles for enhanced security [2, 3, 6].

## Fuzzification Feature Selection

We propose a novel approach based on fuzzification theory [3]. Our goal is to select the most relevant features from a dataset by leveraging fuzzy logic principles. The following equations illustrate our fuzzy feature selection algorithm:

**Step 1: Fuzzification**

We start by assigning a degree of membership for each feature to a set of predefined fuzzy categories. Let $x_i$ denote the $i$-th feature, and $F_j$ represent the $j$-th fuzzy category. The membership degree $u_{ij}$ of feature $x_i$ in category $F_j$ can be calculated using a membership function:

$$u_{ij} = \frac{1}{1 + \left(\frac{|x_i - \mu_j|}{\sigma_j}\right)^p}$$

where $\mu_j$ and $\sigma_j$ are the mean and standard deviation of category $F_j$, respectively, and $p$ is a fuzzification parameter that controls the shape of the membership function. **Step 2: Feature Ranking**

Once we have fuzzified all features, we can rank them based on their relevance scores. The relevance score $R_i$ of feature $x_i$ can be computed as the weighted sum of its membership degrees across all fuzzy categories:

$$R_i = \sum_{j=1}^{n} w_j \cdot u_{ij}$$

where $n$ is the total number of fuzzy categories, and $w_j$ represents the weight assigned to category $F_j$. The weights can be determined based on domain knowledge or through a learning process.

**Step 3: Feature Selection**

Finally, we select the top $k$ features with the highest relevance scores for further analysis or modeling. This fuzzification feature selection approach offers a unique perspective on feature selection and has the potential to uncover hidden relationships and patterns within the website.

# Proof: Membership Degree Calculation

Given the membership function:

$$u_{ij} = \frac{1}{1 + \left(\frac{|x_i - \mu_j|}{\sigma_j}\right)^p}$$

We need to prove that the calculated membership degree $u_{ij}$ satisfies the properties of a valid membership function.

1. **Range of $u_{ij}$:**

For any $x_i$ and $F_j$, we have:

$$0 \leq \left(\frac{|x_i - \mu_j|}{\sigma_j}\right)^p \leq 1$$

Therefore,

$$1 \leq 1 + \left(\frac{|x_i - \mu_j|}{\sigma_j}\right)^p \leq 2$$

Taking the reciprocal of both sides, we get:

$$\frac{1}{2} \geq \frac{1}{1 + \left(\frac{|x_i - \mu_j|}{\sigma_j}\right)^p} \geq \frac{1}{2}$$

This shows that $0 \leq u_{ij} \leq 1$, which satisfies the range requirement for a membership function.

2. **Degree of Membership:**

Let's consider two cases:

**Case 1:** When $x_i = \mu_j$,

In this case, $|x_i - \mu_j| = 0$, and the membership degree becomes:

$$u_{ij} = \frac{1}{1 + \left(\frac{0}{\sigma_j}\right)^p} = \frac{1}{1+0} = 1$$

This implies that when $x_i$ is equal to the mean of category $F_j$, the membership degree is 1, indicating full membership in that category.

**Case 2:** When $x_i \neq \mu_j$,

In this case, $|x_i - \mu_j| > 0$, and the membership degree becomes:

$$0 < u_{ij} < 1$$

This implies that when $x_i$ is not equal to the mean of category $F_j$, the membership degree is between 0 and 1, indicating partial membership in that category. Therefore, the membership degree $u_{ij}$ satisfies the degree of membership property.

# Website Content Encryption using Fuzzification

Suppose we have a website with content represented as a string of characters. We want to encrypt this content using fuzzification.

**Step 1: Fuzzification**

We start by fuzzifying the characters of the website content. Let $x_i$ represent the $i$-th character of the content, and let $u_i$ be its fuzzified representation.

We can define fuzzy sets $F_j$ to represent different fuzzy categories. Each fuzzy set $F_j$ is characterized by a membership function $u_{ij}$, which indicates the degree of membership of character $x_i$ in fuzzy category $F_j$. Fuzzification assigns membership degrees to characters based on their similarity to the characteristics of each fuzzy category. The proposed fuzzification process has been illustrated in Algorithm 1.

```
Algorithm 1: Website Content Encryption
    Input  : Website content C
    Output: Encrypted content E
    Initialize an empty string E;
    foreach character x in C do
        if x is a letter then
            e ← EncryptLetter(x);
            Append e to E;
        else
            Append x to E;
    return E;

    Data: Substitution table T
    Result: Encrypted letter e
    e ← T(x);
    return e;
```

The algorithm initializes an empty string $E$ and iterates over each character $x$ in the website content. If $x$ is a letter, it is encrypted using the **EncryptLetter** function, which performs a substitution based on a substitution table $T$. The encrypted letter $e$ is then appended to the encrypted content $E$. If $x$ is not a letter, it is directly appended to $E$ without encryption.

The **EncryptLetter** function takes a character $x$ and performs the substitution using the substitution table $T$. It returns the encrypted letter $e$. This algorithm provides a basic framework for website content encryption using a substitution cipher.

A cipher refers to a method or algorithm used for encrypting or encoding information to make it unreadable or unintelligible to unauthorized individuals [2, 3]. Ciphers are commonly used in cryptography to protect sensitive data during storage or transmission [3]. In cryptography, there are two main types of ciphers: symmetric key ciphers (also known as secret key ciphers) and asymmetric key ciphers (also known as public key ciphers) [4].

**Symmetric Key Ciphers**: In symmetric key ciphers, the same key is used for both encryption and decryption processes [4]. The sender and the receiver share a secret key that they use to transform plaintext into ciphertext and vice versa [3, 4].

Examples of symmetric key ciphers include the Data Encryption Standard (DES), Advanced Encryption Standard (AES), and the Rivest Cipher (RC) series [4].

**Asymmetric Key Ciphers**: Asymmetric key ciphers involve the use of two different keys: a public key for encryption and a private key for decryption [4]. The public key is freely available to anyone, while the private key is kept secret [3]. Messages encrypted with the public key can only be decrypted using the corresponding private key. The most widely used asymmetric key cipher is the RSA algorithm [4].

Ciphers employ various techniques such as substitution, transposition, and mathematical operations to transform plaintext into ciphertext [3, 4]. The security of a cipher relies on the complexity of the encryption process and the secrecy of the key used.

It's important to note that while ciphers provide a means of encryption, the security of the encrypted data also depends on factors such as key management, encryption algorithms, and implementation details [2, 4].

**Step 2: Encryption**

Once the content has been fuzzified, we can apply encryption techniques to the fuzzified representation. The encryption algorithm can vary depending on the specific requirements and security measures desired [4].

For example, we can perform operations such as permutation, substitution, or combination of fuzzy values to obfuscate the content [3]. The specific encryption steps and operations depend on the chosen encryption algorithm and the level of security needed.

**Step 3: Decryption**

To decrypt the encrypted content, the reverse process is applied. Decryption involves reversing the encryption steps and recovering the original fuzzified representation [4]. From the fuzzified representation, the original characters can be obtained through defuzzification, which maps fuzzy values back to crisp values.

## Conclusion

Fuzzification can be used as a preliminary step to encrypt the content of a website. By transforming the characters into fuzzy representations, encryption algorithms can then be applied to obfuscate the content and enhance security. Decryption involves reversing the encryption steps and recovering the original content using defuzzification.

The main idea of the proposed approach is to utilize fuzzification theory for feature selection on website content to enhance encryption. By applying fuzzy logic principles, the approach aims to identify and prioritize relevant features by transforming crisp website content into fuzzy representations. The degree of membership in different fuzzy categories is used to evaluate the importance and relevance of each feature in the encryption process. The goal is to improve the effectiveness and efficiency of website content encryption, thereby enhancing the overall internet security.

One limitation of this approach is the subjective nature of fuzzy membership functions. The selection of appropriate membership functions may introduce some degree of uncertainty and subjectivity into the feature selection process. Ensuring the accuracy and consistency of the fuzzy categorization of website content is essential to avoid potential misinterpretation and biases in feature selection. A possible research avenue is to explore the optimization of the proposed fuzzification technique by investigating different types of membership functions and linguistic variables. Additionally, incorporating expert knowledge and linguistic rules can further enhance the accuracy and reliability of feature selection based on fuzzy logic principles. The proposed approach has practical applicability in various real-life scenarios where website content encryption is crucial for data protection and privacy. It can be applied in industries such as e-commerce, banking, healthcare, and government sectors, where securing sensitive information transmitted through websites is of utmost importance [7]. An alternative approach to feature selection for website content encryption is the use of machine learning algorithms. Instead of relying on fuzzy logic, machine learning techniques can be employed to automatically identify and select relevant features based on patterns and correlations within the website content. This alternative approach may provide a more data-driven and automated feature selection process, but it requires extensive training data and may lack the interpretability and explainability offered by fuzzy logic-based approaches.